\pdfoutput=1
\documentclass[doublecol]{epl2} 
\usepackage{graphicx, calc, bm, epsfig}

\title{Effect of time delay on feedback control of a flashing ratchet}
\shorttitle{Delayed feedback control of ratchet}

\author{E. M. Craig\inst{1},  B. R. Long\inst{1}, J. M. R. Parrondo\inst{2} \and H. Linke\inst{1}}
\shortauthor{E. M. Craig \etal}

\institute{                    
  \inst{1} Department of Physics, Materials Science Institute, University of Oregon - Eugene, OR USA 97403\\
  \inst{2} Dep. de F{\'i}sica At{\'o}mica, Molecular y Nuclear and GISC, Universidad Complutense de Madrid, 28040-Madrid, Spain 
}

\abstract{It was recently shown that the use of feedback control can improve the performance of a flashing ratchet. We investigate the effect of a time delay in the implementation of feedback control in a closed-loop collective flashing ratchet, using Langevin dynamics simulations. Surprisingly, for a large ensemble, a well-chosen delay time improves the ratchet performance by allowing the system to synchronize into a quasi-periodic stable mode of oscillation that reproduces the optimal average velocity for a periodically flashing ratchet. For a small ensemble, on the other hand, finite delay times significantly reduce the benefit of feedback control for the time-averaged velocity, because the relevance of information decays on a time scale set by the diffusion time of the particles. Based on these results, we establish that experimental use of feedback control is realistic.}

\begin{document}

\maketitle

\section{Introduction\label{intro}}

Flashing ratchets rectify the thermal motion of diffusive particles by exposing them to a time-dependent, spatially periodic and asymmetric potential \cite{bug, ajdari, astumian, reimann, prost}. These systems are attracting significant interest \cite{linke} because they may be applied, for example, in particle separation \cite{bader1,bader2} or as a power source for polymer motors \cite{downton}. In addition, flashing ratchets are one of the simplest realizations of Brownian motors in general \cite{reimann}, and are therefore of fundamental interest in non-equilibrium statistical mechanics and as models for synthetic \cite{kay} or biological \cite{astumian2} molecular motors. 

In most studies of flashing ratchets, the potential is switched periodically or randomly, without regard to the particle distribution. Recently, a flashing ratchet with feedback control was introduced, where the external potential depends on the state of the system \cite{cao}. The instantaneous center-of-mass velocity was maximized by turning the potential on only when the ensemble-averaged force exerted by the potential would be positive. This strategy maximizes the time-averaged center-of-mass velocity for one particle ($N = 1$) and performs better than a periodically flashing ratchet for $N < 10^2 -10^3$. However, because fluctuations are needed to trigger the next switching event, the time-averaged velocity goes to zero for large $N$ where center-of-mass fluctuations become rare. In an improved feedback strategy \cite{dinis}, the potential is switched on (off) whenever the average force increases (decreases) through set thresholds, eliminating the need to wait for fluctuations. This protocol performs as well as the feedback strategy in \cite{cao} for $N = 1$, and yields approximately the same current as optimal periodic flashing for large $N$. 

In an experiment designed to implement such feedback strategies, there will be a finite time lag between the collection of information and any feedback to the system because of the time it takes to acquire an image of particles, and to determine the particle positions and ensemble averages. Delayed feedback has been demonstrated to produce complicated dynamics in chaotic, inertial ratchets \cite{kostur, son, wu}, stochastic systems that display a Hopf bifurcation (oscillatory instability) in the absence of delay \cite{wolfrum} and biological systems \cite{mackey, may, tyson}, and some analytical methods have recently been developed to study time delay in stochastic systems \cite{frank, guillouzic, gu}. Here we study the impact of time delay on the effectiveness of the feedback control strategy introduced in \cite{cao}. 

For a small number of particles (less than $N \approx 300$), we find that finite delay times in the feedback significantly reduce the average velocity, because the quality of the collected information decays on the scale of the system's characteristic diffusion time. This is consistent with the result \cite{cao2} that loss of information about the system limits the improvement of the flux for the feedback ratchet strategy in \cite{cao} over a periodically flashing ratchet. On the other hand, for the quasi-deterministic case of  $N  > 10^2 - 10^3$, we find that a well-chosen delay time effectively reproduces the advantageous threshold protocol in \cite{dinis} even though no threshold criterion is actually used in the algorithm. As a result, a time delay in the control strategy of \cite{cao} surprisingly can increase the average current. Somewhat paradoxically, the system is found not only to tolerate, but actually to benefit from the experimental ``handicap" of finite time lags. 

\section{Model\label{model}} 

\begin{figure}
\onefigure{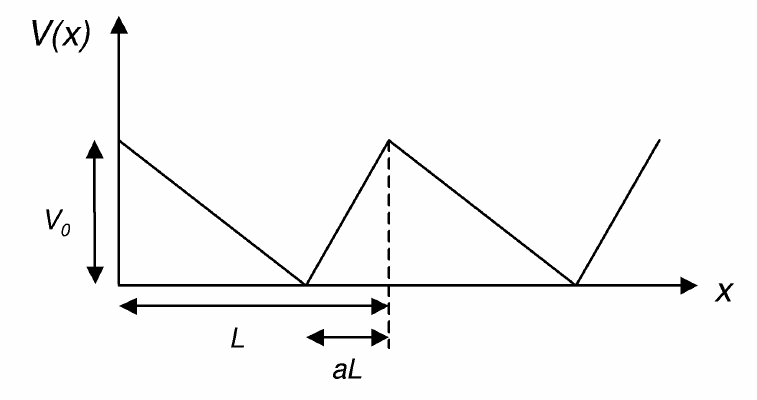}
\caption{The applied ratchet potential $V(x)$, shown in the schematic above, is characterized by periodic length $L$, height $V_0$, and asymmetry $a$.}
\label{fig1}
\end{figure}

We consider the following one-dimensional system discussed in \emph{Cao et. al.} \cite{cao}: An ensemble of $N$ non-interacting, overdamped particles is exposed to a spatially periodic,  piecewise linear `sawtooth' potential $V(x)$, characterized by the spatial period, $L$, height $V_0$, and asymmetry $a$ (Fig.~\ref{fig1}). The motion of the particles is described by the Langevin equations:

\begin{equation}
\label{eq:lang}
\gamma \dot{x}_i(t) = \alpha (t) F(x_i(t)) + \xi_i(t); \mbox{\hspace{0.5 in} i = 1,\ldots, N}
\end{equation}
where ${x}_i(t)$ is the position of particle $i$, $\gamma$ is the drag coefficient of a particle, and $\xi_i(t)$ is a randomly fluctuating Gaussian white noise term with zero mean and correlation $\left<\xi_i(t) \xi_j (t^{ \prime}) \right>=2\gamma kT \delta_{ij} \delta (t-t^{ \prime})$. The external force is given by $F(x) = -V^{ \prime} (x)$, and $\alpha(t)$ is a control parameter which can take the value of 1 or 0, thus switching the external potential on or off.  

In \cite{cao}, the following two control strategies are compared:

(1) Periodic switching: $\alpha(t+ \tau) = \alpha(t)$, with $\alpha(t) = 1$ for $t \in [0, \tau/2)$, and $\alpha(t) = 0$ for $t \in [\tau/2,\tau)$. 

(2) Controlled switching (maximization of the instantaneous center-of-mass velocity):

\begin{equation}
\label{eq:alpha}
\alpha(t) = \Theta(f(t)),
\end{equation}

where

\begin{equation}
\label{eq:force}
f(t) = \frac{1}{N} \sum_i^N F(x_i)
\end{equation}
is the ensemble average of the force the particles would feel if the potential were on, and $\Theta(y)$ is the Heaviside function, $\Theta(y) = 1$ if $y \geq 0$ and 0 otherwise.

Here, we consider two distinct types of time delay in the feedback control (Fig.~\ref{fig2}):

(1) Implementation time, $t_1$: If a measurement is taken at time $t$, any feedback (i.e. a possible change in potential) based on this measurement will occur at time $t + t_1$. This type of delay is caused by data processing in an experiment. 

(2) Measurement interval, $t_2$: If a measurement is taken at time $t$, the next measurement will be taken at time $t + t_2$. This type of delay could be due to a limit in the readout rate of a camera system which acquires images of the particle distribution. 

We use Langevin dynamics simulations to study the effect of these time delay parameters on the average particle velocity achieved by the control strategy described by eq.~\ref{eq:alpha}. In the presence of time delay, the motion of the particles is given by
\begin{equation}
\label{eq:lang2}
\gamma \dot{x}_i(t) = \beta (t) F(x_i(t)) + \xi_i(t); \mbox{\hspace{0.5 in} i = 1,\ldots, N}
\end{equation}
where $\beta(t)$ is the actual state of the system based on a delayed response to the cue $\alpha(t)$. In all simulations, $kT=1$, $kT / \gamma = D = 1$, $L = 1$, and $a = 1/3$. We will use $V_0=5kT$ unless otherwise stated. The finite time step, $dt$, in a Langevin dynamics calculation introduces an unintended measurement delay $t_2$, but by choosing a sufficiently small $dt$, we can examine the behavior of the system in the $t_2=0$ limit (for example, for $N=1$, we use $dt=10^{-6} L^2/D$). To measure ratchet performance, we calculate the time-averaged center-of-mass velocity, $v_{CM}= \langle \dot x_{CM} \rangle$. We have tested that $v_{CM}$ does not depend on the initial distribution, when averaged over a long enough period of time. 

\section{Results\label{results}}

We begin by considering the effect of implementation delay, $t_1$, with zero measurement interval ($t_2 = 0$), and show in  fig.~\ref{fig5} $v_{CM}$ as a function of  $t_1$ for different $N$. The impact of implementation delay on feedback effectiveness depends on the ensemble size as follows:

\textbf{Small $N$:} For $N=1$ and perfect feedback ($t_1 = 0$), we reproduce the result $v_{CM} = 4.27$ $D/L$ \cite{cao}, an order of magnitude faster than the velocity $v_{opt}  \approx 0.284 D / L$ for optimal periodic flashing without feedback for all $N$ ($\tau = \tau_{opt} \approx 0.1 L^2 / D$).  However, the velocity decreases very quickly with $t_1$: for $N<10^3$, $v_{CM}$ drops below $v_{opt}$ for delay greater than $t_1 \approx 0.03 -0.05 L^2/D$ (fig.~\ref{fig5}). This time scale is reasonable when one considers that the distance, $\sqrt{2Dt}$, a particle diffuses during $t=0.05 L^2 / D$ is about $0.3 L$, comparable to the ratchet's critical length scale of $aL$. In other words, for small $N$, the implementation delay at which the feedback strategy loses effectiveness is comparable to the correlation time between measured and actual particle positions with respect to the potential features.

\textbf{Large $N$; small $t_1$:} As the number of particles increases, center-of-mass fluctuations become more rare, and the time-evolution of the ensemble becomes more deterministic. Thus, for larger ensembles, the effect of a short implementation delay $t_1$ can be visualized by inspecting the characteristic time development of $f(t)$ for $N=10^6$ (fig.~\ref{fig3}). Initially, the potential is on and the particles are equilibrated, so that $f \approx 0$. A fluctuation that yields $f  < 0$ produces a cue to turn the potential off (setting $\alpha(t)$ to zero), which is implemented a time $t_1$ later ($\beta (t)= \alpha(t-t_1)$ for $t_2=0$). This leads to a rapid decrease of $f$ because half of the particles quickly diffuse into an area with a large negative force. Rapid shifts in $f$ occur for a duration of about $t_1$, as $\beta(t)$ switches in response to previous cues. Once $\beta = 0$ becomes steady, the system approaches its new equilibrium state. Near the new equilibrium, a center-of-mass fluctuation yields $f  > 0$, and after a delay of $t_1$ the potential is turned on, rapidly pushing the particles down the potential slopes (the steeper slope is cleared first, yielding an initial increase of $f$). Again, fluctuations in $f$ are observed for a duration of about $t_1$ before $\beta=1$ becomes steady. For small $t_1$, the velocity is reduced due to the additional time spent near equilibrium before a cue is implemented and the unproductive fluctuations in $f$ after the potential is switched.

\textbf{Large $N$; large $t_1$:} For large $N$, we find that a finite value of $t_1$ can be chosen such that $v_{CM}$ surprisingly is larger than in the absence of time delay. This result is initially unexpected because for $N > 10^3$, $v_{CM}$ achieved by the feedback mechanism used here (eq.~\ref{eq:alpha}) falls short of $v_{opt}$, even without the experimental handicap of finite implementation delay \cite{cao}. To understand how implementation delay can actually improve the performance of a feedback mechanism, note that $f(t)$ essentially becomes deterministic for large $N$. As shown in fig.~\ref{fig3}(b), for $t_1 = 0.09 L^2 / D $, after a small number of initial cycles, a cue created in one cycle of the potential correctly triggers a switching event in the \emph{next} cycle. This mode is stable when $t_1$ is larger than the duration of the initial transient toward the maximum $|f|$ that is observed after each switching event, because a potential switch after the system moves through the maximum $|f|$ will trigger a sign change of $f$, and thus a new cue. For the parameters used here, the maximum $|f|$ is reached after about $0.03 L^2/D$, which is consistent with our observation that $v_{CM}$ increases with $t_1$ for $N  > 10^3$ when $t_1> 0.03 L^2/D$. 

A key result of this paper is that, for a large enough ensemble, there is a range of implementation delays that improve the performance of a feedback-controlled ratchet by allowing the system to synchronize into a stable mode of oscillation that does not rely on fluctuations to trigger a change in the potential. If $t_1$ is longer than the initial approach to the maximum $|f|$ in the characteristic time-evolution of $f$ ($t_1>0.03L^2/D$), then a cue created in one cycle can trigger a switching event in the next cycle. This produces a stable mode of oscillation with a period of about $\tau = t_1$ (fig.~\ref{fig4}(b)). Thus, for $t_1 \approx \tau_{opt}  = 0.1L^2/D$, the system operates at $v_{opt}$ (Fig.~\ref{fig4}(a)). The inset of fig.~\ref{fig4}(a) shows that this behavior is qualitatively independent of $V_0$, noting that $v_{opt}$ is itself a function of $V_0$. For increasing $V_0$, $v_{opt}(V_0)$ is reached for smaller $t_1$, because less time is required for the particles to localize in the potential, decreasing the implementation delay necessary to achieve $v_{CM}=v_{opt}(V_0)$.

Interestingly, for even larger $t_1$ (we investigated up to $t_1=0.5L^2/D$), the system usually finds a mode in which cues are implemented multiple periods in the future. In this case, it can take hundreds of cycles for the system to stabilize in this mode of operation. As seen in fig.~\ref{fig4}(b), for values of $t_1$ such that a cue is implemented after two cycles (at around $t_1> 0.16 L^2/D$), the stable period is $\tau \approx t_1/2$. Near the boundary between two modes, the system can synchronize into either mode. In the range of $t_1=0.13 - 0.18 L^2/D$, the system has been observed to become metastably synchronized into one quasi-periodic mode for several hundred cycles before locking into the other mode (for example, see inset to fig.~\ref{fig4}(b)). For $N=10^6$, we observe these stable modes for at least $10^4$ cycles, and one can expect that periodic modes of oscillation will always remain stable in the deterministic limit of large $N$. 

\textbf{Measurement delay:} The synchronized behavior discussed above relies on the arrival of one cue triggering a new cue of the same sign. We now examine the impact of a finite measurement interval, $t_2$, on this synchronization, for several values of $t_2$ relative to $t_1$. For $N = 10^6$ (fig.~\ref{fig6}(a)), broadly speaking, we observe two regimes of behavior: (1) For $t_2 \leq t_1$, we see qualitatively the same behavior as for $t_2=0$, except that the velocity versus $t_1$ behavior shifts to smaller $t_1$, because increasing $t_2$ increases the period of the synchronized cycles. If $t_2$ is not close to being an integer multiple of $t_1$, for example $t_2=t_1/2$, $v_{opt}$ is not achieved because a new measurement is taken before the previous measurement has been implemented, which interferes with the synchronization necessary to produce optimized periodic flashing. (2) For $t_2>t_1$, a new cue is sent after the force has switched signs in response to the previous cue, so the synchronization-induced increase in the average velocity with $t_1$ does not occur.

On the other hand, for small $N$, the velocity decreases with $t_1$ for any value of $t_2$ (shown for $N=10$ in fig.~\ref{fig6}(b)), because increasing $t_1$ decreases the level of correlation between the particle distribution when the cue is taken and when it is implemented. Interestingly, for $0<t_2<t_1$, the falloff in velocity with respect to $t_1$ is not as steep as when $t_2=0$ for the following reason: A finite measurement interval $t_2$ allows the ensemble to evolve before the next cue is taken, which decreases the amount of unproductive rapid oscillations. For $t_2>t_1$, the velocity is slower than when $t_2=0$, because there is idle time in each cycle before a new measurement is taken.

\section{Experimental feasibility}

\begin{figure}
\onefigure{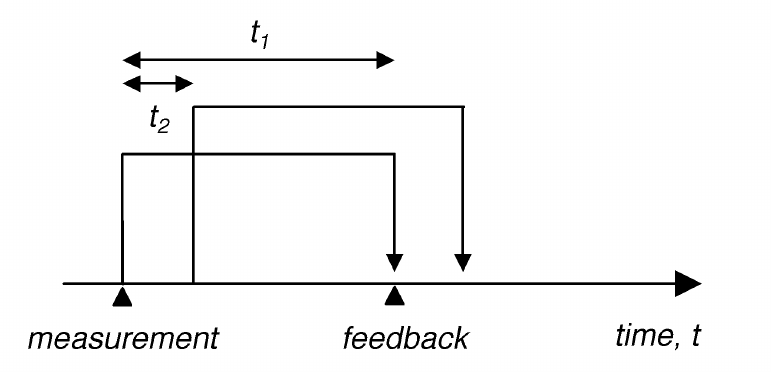}
\caption{The schematic above illustrates the two types of time delay in feedback strategies: If a measurement is taken at time $t$, the feedback based on this measurement will be implemented at time $t+t_1$ (implementation delay). The next measurement will be taken at time $t + t_2$ (measurement interval).}
\label{fig2}
\end{figure}

\begin{figure}
\onefigure[width=\columnwidth]{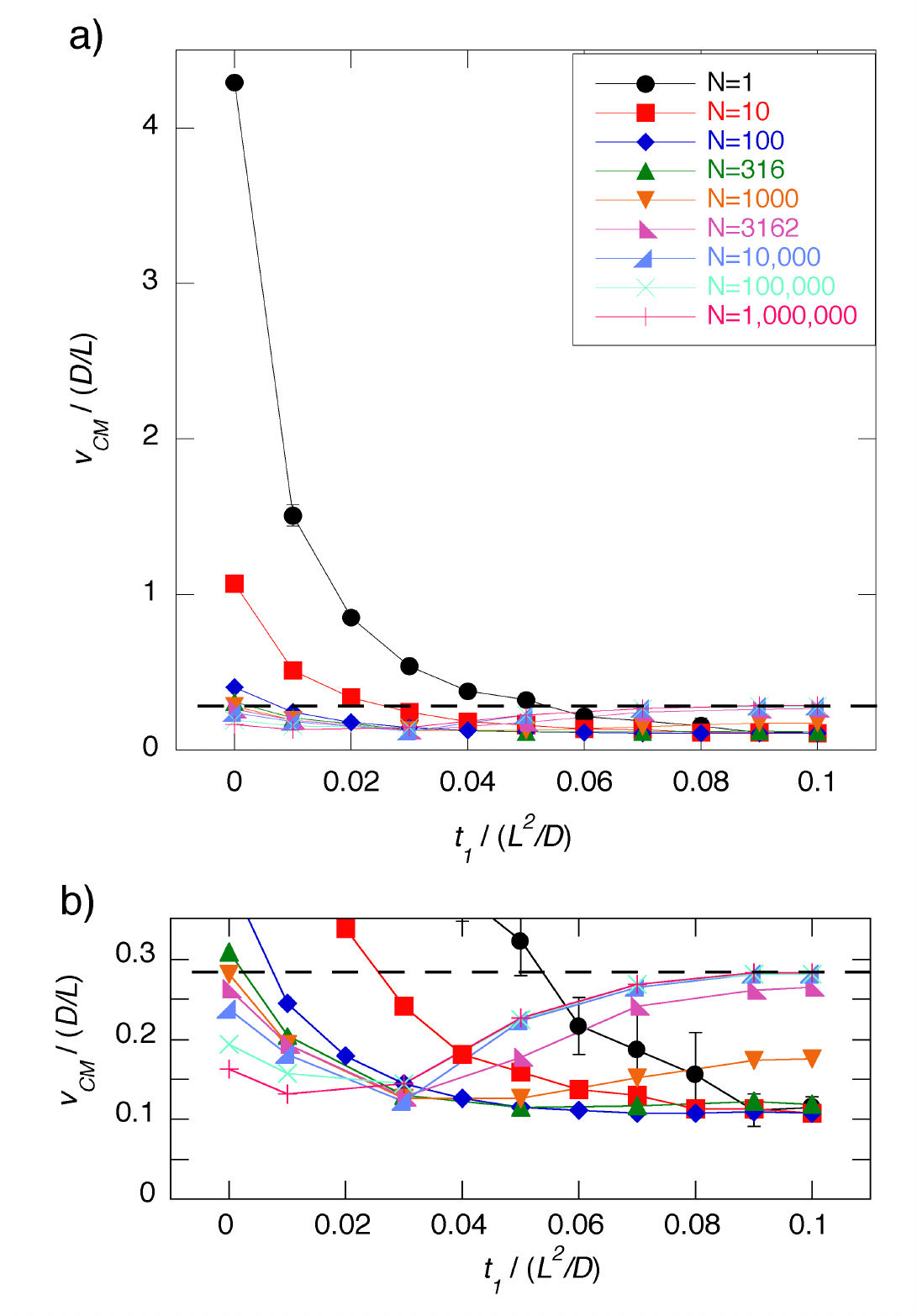}
\caption{Langevin dynamics simulations. (a) Average velocity as a function of $t_1$, for $t_2 = 0$ and different $N$ as indicated. The lines between data points are included as a guide to the eye. The horizontal, dashed line shows the average velocity for switching with optimal period $\tau_{opt}= 0.1 L^2 / D$. (b) An enlarged view of (a) for small $v_{CM}$. Error bars are shown for $N=1$. For $N \ge 10$, the error is smaller than $0.1 \%$ of the values of $v_{CM}$ displayed.}  
\label{fig5}
\end{figure}

\begin{figure}
\onefigure[width=\columnwidth]{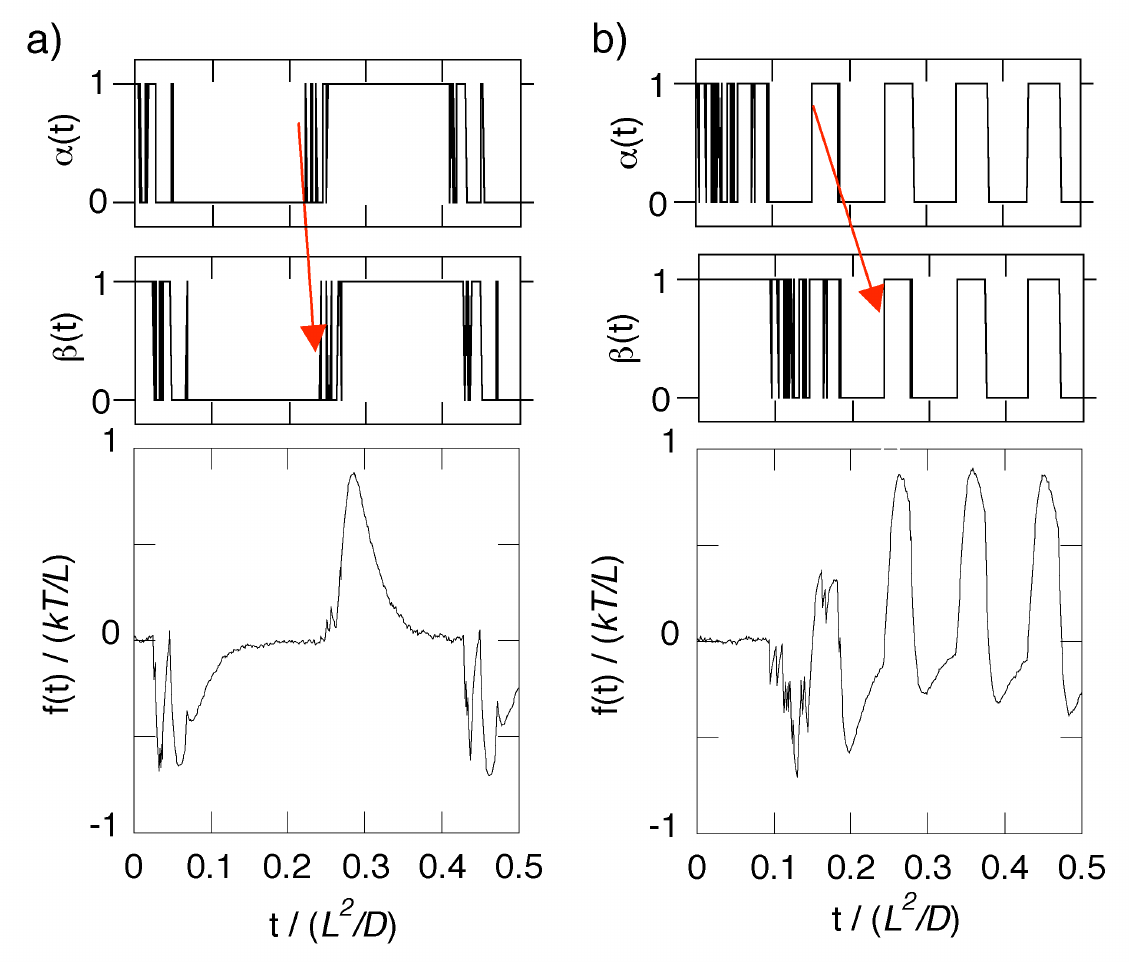}
\caption{Langevin dynamics simulations. (a) Average force $f(t)$ for $N = 10^6$, $t_1 = 0.02 L^2 / D$, and $t_2 = 0$. The measured state, $\alpha(t)$ (cue) and the implemented state, $\beta(t)$ (response) are shown above. The red arrows indicate the delay between cue and response. Initially, the potential is on, and particles are equilibrated in this potential. (b) $N = 10^6$, $t_1 = 0.09 L^2 / D$, and $t_2 = 0$. }   
\label{fig3}
\end{figure}

\begin{figure}
\onefigure[width=\columnwidth]{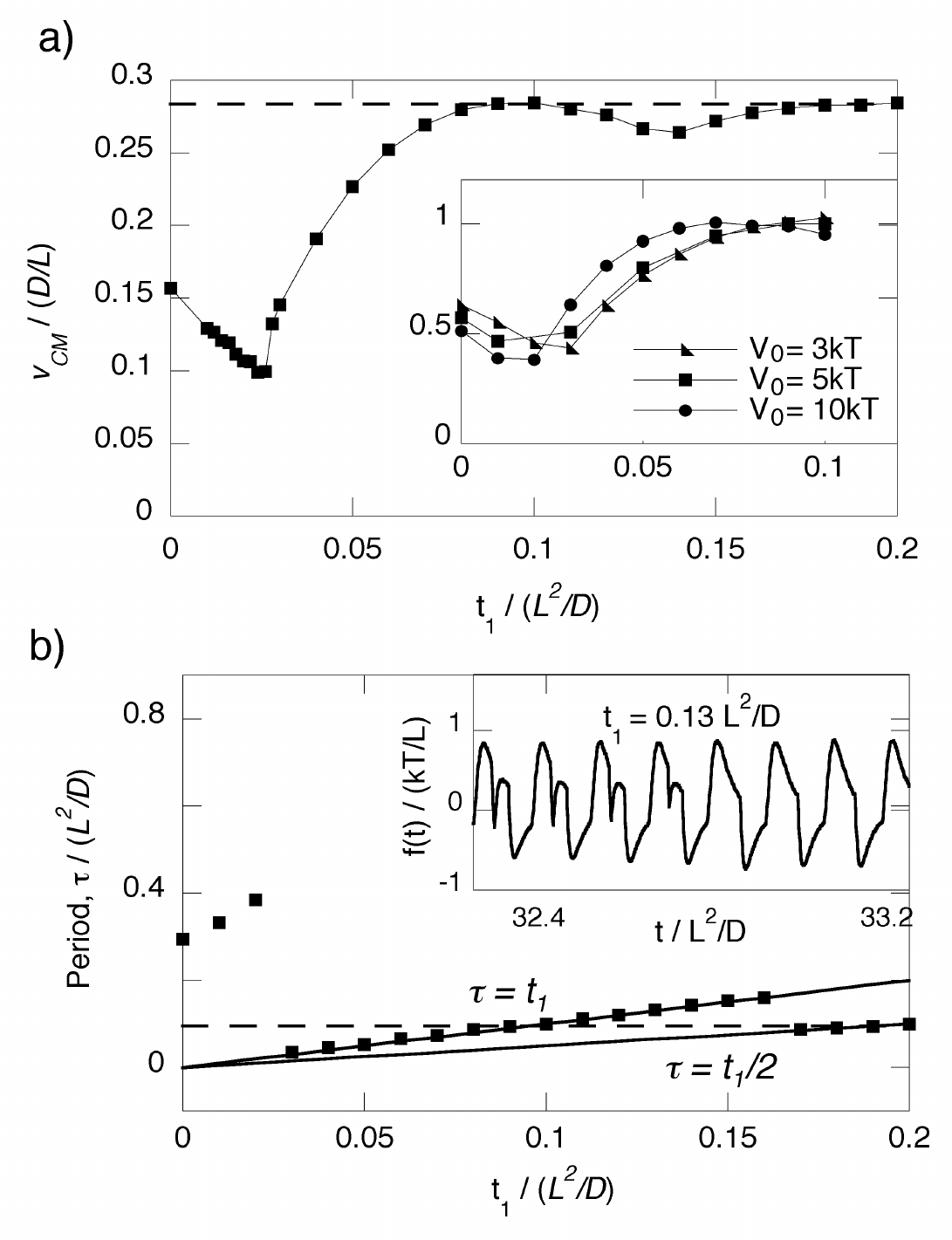}
\caption{Langevin dynamics simulations ($N=10^6$ data from fig.~\ref{fig5}). (a) Average velocity as a function of $t_1$, for $t_2 = 0$ and $N = 10^6$. The lines between data points are included as a guide to the eye. The dashed line shows the average velocity, $v_{opt}$, for optimal periodic switching. Inset: Average velocity as a function of $t_1$ for three different $V_0$, each normalized by the $v_{opt}$ corresponding to that $V_0$ with $a=1/3$. (b) Average temporal period of stable oscillations as a function of $t_1$, calculated using the same simulated data as in (a). The dashed line corresponds to $\tau_{opt}$. The slope of lines through the data points are characteristic of different modes of quasi-periodic stable oscillations. Data points for $t_1 > 0.03 L^2/D$ either follow $\tau = t_1$, which means that a cue created in one cycle triggers a switching event in the next cycle, or $\tau = t_1/2$, which means that a cue in one cycle is implemented after two cycles. Inset: Example of a transition from a metastable to a stable mode, for $t_1=0.13L^2/D$.}   
\label{fig4}
\end{figure}

\begin{figure}
\onefigure[width=\columnwidth]{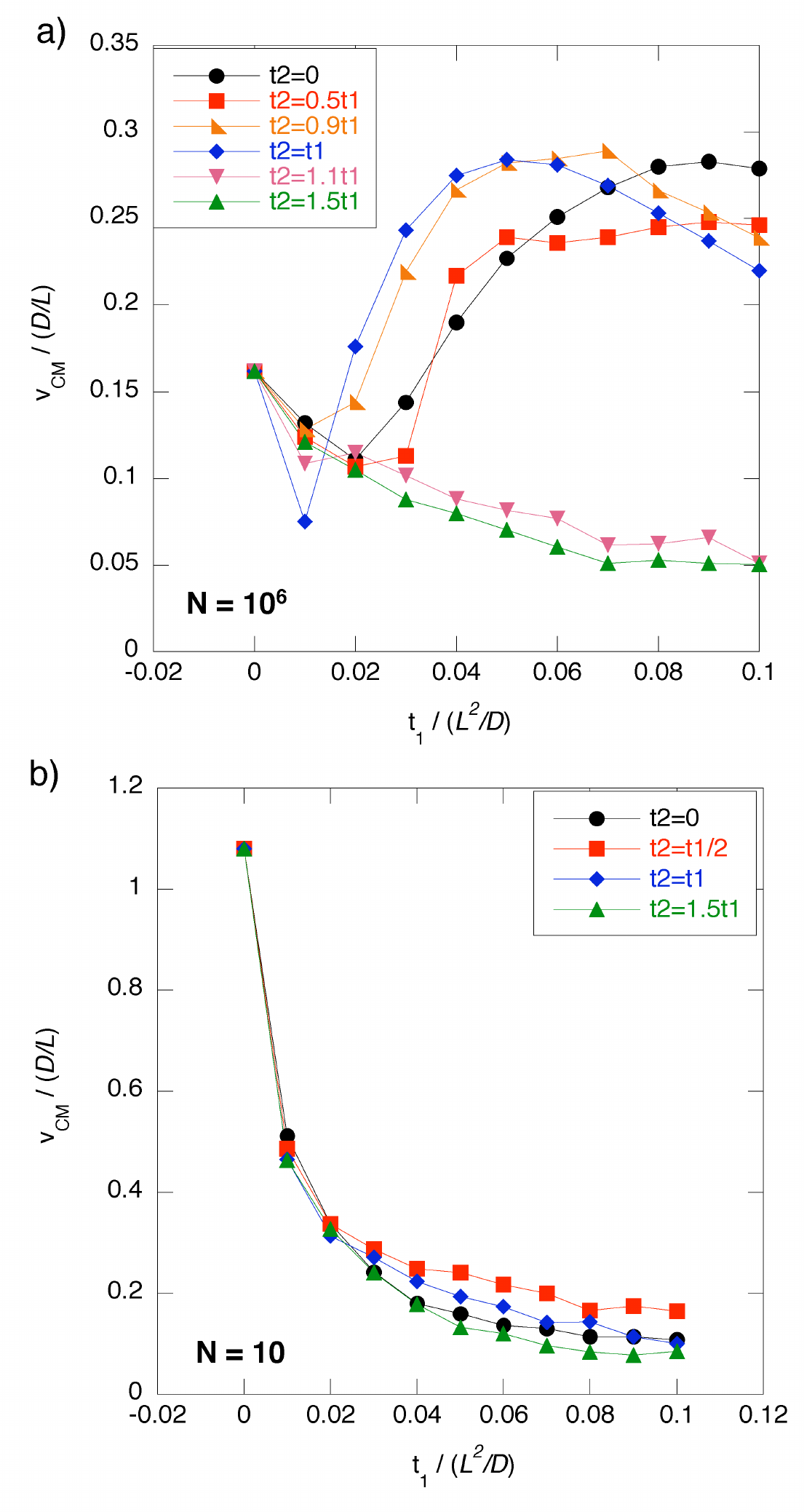}
\caption{Langevin dynamics simulations. (a) Average velocity as a function of $t_1$ for $N = 10^6$. There are two regimes with qualitatively different behavior, $t_2<t_1$ and $t_2>t_1$, as discussed in the text. (b) Average velocity as a function of $t_1$ for $N = 10$, for several choices of $t_2/t_1$. The lines between data points are included as a guide to the eye.}
\label{fig6}
\end{figure}

To place the time scales discussed here into a quantitative context, consider a possible experimental implementation that uses an interdigitated electrode array to expose charged, colloidal particles such as carboxylate-modified, fluorescent polystyrene beads to a time-dependent, asymmetric, electrostatic potential. Such flashing ratchet devices have been used previously with DNA molecules \cite{bader1}. A feedback operation in such a system will require: (I) acquisition of a 2D image using a CCD camera; (II) numerical determination of particle locations; (III) decision whether to switch the potential; and (IV) change of voltage applied across the array. The total time for operations  (I) - (IV) corresponds to $t_1$ in fig.~\ref{fig2}. The camera exposure time (step I) often limits the rate at which images can be acquired and corresponds roughly to $t_2$ in fig.~\ref{fig2} \footnote{The correspondence between exposure time and $t_2$ as defined in fig.~\ref{fig2} is not exact because, in an experiment, particle positions are averaged over the exposure time, whereas in our numerical studies particle positions are collected instantaneously at the end of $t_2$. Nevertheless, our general discussion of relevant time scales holds true.}. 

For small $N$, fig.~\ref{fig5} shows that to fully benefit from feedback, delay times should be kept below about $0.01 L^2/D$. Feedback schemes to control the position of a single, fluorescent particle have been implemented with rates up to 300 Hz \cite{cohen}. If we assume $t_1 \approx 10ms$ for small $N$, then $t_1< 0.01 L^2/D$ when $L > 1.2 \mu m$ for beads of diameter $d = 300 nm$ in an aqueous environment (viscosity $\eta = 10^{-3} Pa \cdot s$, $D = kT/3 \pi \eta d$). In other words, for $N \approx 1$, a ratchet with $L \approx 1 \mu m$ or larger could yield almost the full theoretical benefit of feedback. Electrodes of this size can easily be fabricated using standard lithography.

The rates quoted above are accessible when a small subset of the CCDÕs pixels are used.  For $N \approx 10^2 - 10^3$, a full-size (for example 512 x 512) image would be needed, yielding a typical exposure and readout time from a high-sensitivity CCD camera of about $30 ms$. Ideally, operations (II)-(IV) are performed while the next image is collected by the camara, such that total delay times of order $t_1 =2 t_2 \approx 60 ms$ are realistic. To observe the effect of feedback delay on ratchet transport shown in Fig.~\ref{fig6}, it is desirable to vary $t_1$ across the range $0.01 - 0.1 L^2/D$. For delay times on the order $60 ms$, one therefore requires $L^2/D \approx 6 s$.  For beads of diameter $d$ in an aqueous environment exposed to a ratchet potential with $L = 3 \mu m$, we find $L^2/D = 6 s$ for $d = 290 nm$. 

The time scale $L^2/D$ scales linearly with $d$ for fixed $L$ and can thus easily be adjusted by changing the particle size. If locating individual particles is either too slow or limited by image resolution, it should be possible to establish the desired feedback protocols by measuring density distributions. The effects demonstrated above numerically are most clearly observable for potential depths of several $kT$ or more. In water, commercially available $300nm$ carboxylate-modified polystyrene beads have a charge $q \approx 10^5 e$, so an electrostatic potential in the $100mV$ range should create a trapping strength $qV$ much greater than $kT$. It therefore appears realistic in principle to experimentally explore the key results presented here.

\section{Concluding remarks\label{conclusion}}

In conclusion, we established that an implementation delay on the order of the diffusion time over the length of the negative-slope portion of the ratchet potential significantly compromises the performance of a feedback-controlled flashing ratchet for small particle numbers (fewer than $\approx 300$). However, we found that for larger particle numbers, an implementation delay actually improves the performance of the feedback strategy introduced by Cao et al. \cite{cao}, and matches the performance of an optimally periodic potential modulation, similar to the feedback mechanism discussed in \cite{dinis}, even for surprisingly long delay times. 

We have demonstrated that this key result is true in general for $t_2 \leq t_1$, independent of the potential depth and the initial conditions. We expect that the transition between the (`nondeterministic') low $N$ behavior and the (`deterministic') high $N$ behavior is weakly dependent on $V_0$. For the relevant case of potential depths of several $kT$ or more, increasing the potential will have minimal effect on how deterministic the system is, and the transition between `low $N$' and `high $N$' regimes will depend most critically on the behavior of the ensemble when the potential is off. We have modeled the role of time delay in feedback control of a ratchet over a range of parameters that would be feasible to implement experimentally. 

\textbf{Note:} After submission, two preprints \cite{caonew1, caonew2} came to our attention, which use a combination of numerical and analytical methods \cite{frank} to study the role of implementation time delay in a flashing ratchet. 

\acknowledgments
We are grateful to Martin Zuckermann and Ben Lopez for helpful discussions. This research is based upon work supported by the National Science Foundation under CAREER Grant No. 0239764, and has been enabled by the use of the Bugaboo computing facility and the IRMACS center of Simon Fraser University, which are funded in part by the Canada Foundation for Innovation. J. M. R. P. acknowledges financial support from Ministerio de Educaci\'on y Ciencia (Spain), grant MOSAICO, and from BCSH, grant UCM PR27/05-13923.

\end{document}